\documentclass[12pt]{article}
\usepackage[dvipsnames]{xcolor}
\usepackage{newtxtext,newtxmath}
\usepackage{graphicx}

\usepackage[letterpaper,margin=1in]{geometry}

\renewenvironment{abstract}
	{\quotation}
	{\endquotation}

\date{}

\makeatletter
\renewcommand{\fnum@figure}{\textbf{Figure \thefigure}}
\renewcommand{\fnum@table}{\textbf{Table \thetable}}
\makeatother

\usepackage{scicite}

\usepackage{url}

\def\scititle{
	Nereid as a Regular Satellite of Neptune
}
\title{\bfseries \boldmath \scititle}

\author{
	Matthew~Belyakov$^{1\ast}$,
	M.~Ryleigh~Davis$^{1}$,
	Ian~Wong$^{2}$,
    Konstantin~Batygin$^{1}$,\and
    Michael~E.~Brown$^{1}$\and
	\small$^{1}$Division of Geological and Planetary Sciences, California Institute of Technology, Pasadena 91125, USA.\and
	\small$^{2}$Space Telescope Science Institute, Baltimore, MD 21218, USA\and
	\small$^\ast$Corresponding author. Email: mattbel@caltech.edu
}

\begin{document} 

% Insert the title and author list
\maketitle

% Abstract, in bold
% There are strict length limits, and not all formats have abstracts.
% Consult the journal instructions to authors for details.
% Do not cite any references in the abstract.
\begin{abstract} \bfseries \boldmath
Nereid, Neptune's third largest moon, is considered to be a captured irregular satellite due to its highly eccentric orbit. However, among irregular satellites, Nereid is an outlier: it is the largest, the closest to its host planet, and the most eccentric. We present James Webb Space Telescope near-infrared spectroscopy of Nereid that demonstrates that its composition is inconsistent with its suggested captured origin. We then simulate Nereid's early orbital history subsequent to Triton's capture to demonstrate a plausible dynamical pathway for a regular satellite formed in-situ around Neptune to evolve to Nereid's present-day orbit. Based upon the available spectroscopic and dynamical evidence, we propose that Nereid is not a body captured from the Kuiper belt, but rather the sole surviving intact regular satellite of Neptune.
\end{abstract}

% \subsection*{Teaser}
% Evidence from James Webb Space Telescope suggests Neptune's third largest moon, Nereid, is the planet's only intact original moon.

% The first paragraph of any Science paper does NOT have a heading
% Nor is it indented
\subsection*{Introduction}
\noindent
Neptune is the only giant planet that lacks an intact regular satellite system. The planet's largest moon, Triton, constitutes over 99\% of the satellite system's mass and follows a 157 degree retrograde, synchronous orbit, at 14 Neptune radii. Triton's unique dynamical state among Solar System moons suggests the satellite was captured from the protoplanetary disk, with its orbit subsequently circularized through tidal dissipation or collisions \cite{Goldreich1989Sci, Tsui2002P&SS, Agnor2006Natur, Nogueira2011Icar}. The conclusion that Triton was captured is further supported by notable similarities between Pluto and Triton \cite{McKinnon1984Natur, mousis_triton}. 

Neptune also hosts a set of seven inner moons often referred to as ring-moons. As suggested by their name, ring-moons share material and an origin with the planet's rings. Neptune's ring-moons are likely the collisionally disrupted remnants of the proto-Neptunian disk \cite{Goldreich1989Sci, Banfield1992Icar, Cuk2005ApJL}. Voygaer imaging of Proteus, Neptune's second largest moon and the largest ring-moon, revealed a rough surface and a shape inconsistent with hydrostatic equilibrium, indicating Proteus may be a rubble pile \cite{Croft1992Icar}. Therefore, the ring-moons do not qualify as intact and original Neptunian satellites.

Nereid is the third largest Neptunian satellite, and is typically classified as an irregular satellite due to its highly eccentric orbit\cite{Kuiper1949PASP}. Irregular satellites are thought to be captured from planetesimals in the protoplanetary disk  \cite{Kuipersat, 2007AJ....133.1962N}. Among the irregular satellites of the giant planets, Nereid has the lowest pericenter (0.012 Hill radii or 56 Neptune radii) and semi-major axis (0.048 Hill radii or 224 Neptune radii), while typical irregular satellites have orbits outside 0.1 Hill radii of their host planets. Nereid also has the highest mean eccentricity ($e\simeq0.75$) and the largest radius at 175~km -- twice the size of Saturn's Phoebe \cite{Thomas1991JGR, Kiss2016MNRAS}. Compositionally, Nereid also stands out: it has a notably higher albedo (0.24) than typical similarly-sized Kuiper belt objects (KBOs) \cite{Kiss2016MNRAS}, and has a bluer spectral slope than Neptune's other irregular satellites \cite{Graykowski2018AJ}. Past ground-based near-infrared spectroscopy found deep water ice absorption features on Nereid \cite{Brown1998ApJL, Brown1999Icar, Sharkey2021PSJ}, however these early studies had a limited compositional picture of outer solar system bodies and could not definitively determine Nereid's origin. Here, we use James Webb Space Telescope observations of Nereid coupled with dynamical simulations to ascertain the origin and evolution of this unique moon.

\subsection*{Results}
\subsubsection*{JWST Spectroscopy}
\begin{figure} % Do NOT use \begin{figure*}
	\centering
	\includegraphics[width=0.9\textwidth]{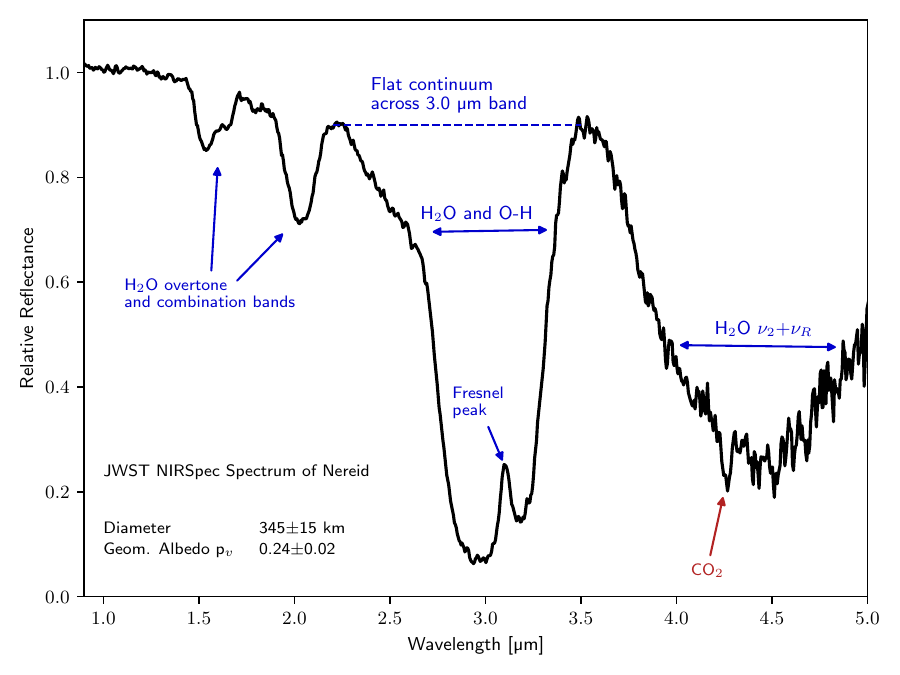} % for an image file named example_figure.*
	% Pick an appropriate width - in print, figures are usually one or two columns wide, which can
	% be approximated by 0.3\textwidth or 0.6\textwidth respectively. Use appropriate label sizes.

	% Captions go below figures
	\caption{\textbf{JWST/NIRSpec spectrum of Nereid.} Relative reflectance spectrum of Nereid with key features highlighted. The near-infrared slope is distinctly negative. Features at 1.5, 2.0, 3.0, and 4.5~{\textmu}m are due to abundant H$_2$O ice, while CO$_2$ produces the 4.27~{\textmu}m band. Past the CO$_2$ absorption feature, high frequency variation in the signal appears owing to lower detector sensitivity, decreasing signal, and increased thermal background.}
	\label{fig:relref} % give each figure a logical label name
\end{figure}

We observed Nereid using the James Webb Space Telescope (JWST) Near-Infrared Spectrograph (NIRSpec) integral field unit (IFU) in Prism mode, which covers 0.6-5.3~{\textmu}m with an R$\sim30-300$ resolving power \cite{supplmat}. Nereid's reflectance spectrum (Fig.~\ref{fig:relref}) is dominated by water ice, with a negative, or blue slope between 1.0 and 2.5~{\textmu}m. The 1.65~{\textmu}m absorption inset on the 1.5~{\textmu}m band, along with the 3.1~{\textmu}m narrow Fresnel peak evidence crystalline water ice \cite{Mastrapa2009ApJ}. JWST's expanded wavelength coverage also reveals the deep 4.5~{\textmu}m H$_2$O combination band. Despite abundant water ice, the continuum across the large H$_2$O band at 3.0~{\textmu}m is flat (dotted line in Fig.~\ref{fig:relref}), unlike both laboratory studies of water ice and outer solar system icy surfaces which have very low reflectances longwards of the 3.0~{\textmu}m water band \cite{Clark1980Icar, Mastrapa2009ApJ, Clark2019Icar}. 

\begin{figure} % Do NOT use \begin{figure*}
	\centering
	\includegraphics[width=0.6\textwidth]{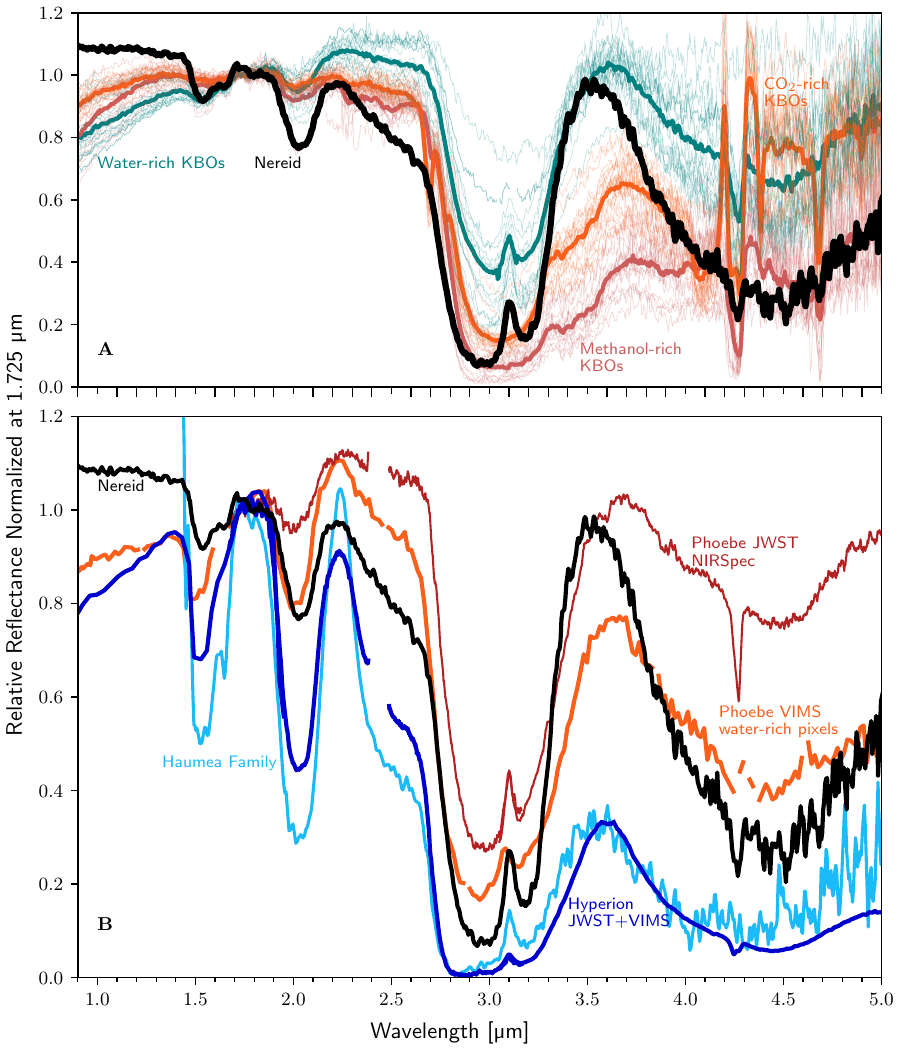} 
	\caption{\textbf{Comparison of Nereid to outer solar system bodies.} (A) Spectrum of Nereid (black) plotted against the three classes of KBOs observed by JWST \cite{Pinilla-Alonso2025NatAs}. The CO$_2$ and methanol-abundant KBOs shown in orange and red, respectively, do not match Nereid which lacks CO or deep 3.4~{\textmu}m absorption features from organics. The water-rich KBOs shown in teal are a better match. However, compared to this population, Nereid has a bluer continuum slope and deeper water ice bands. Nereid's unique spectrum suggests a non-Kuiper belt origin. (B) Nereid's spectrum (black) shown against the JWST Phoebe spectrum in red \cite{Belyakov2025PSJ}, the \textit{Cassini} VIMS water-rich spectra of Phoebe in orange \cite{Fraser2018AJ}, the combined JWST and \textit{Cassini} VIMS global spectrum of Hyperion in blue \cite{Clark2012Icar, Brown2025PSJ}, and the JWST average spectrum of the Haumea family members. Phoebe is a captured water-rich KBO with abundant water ice exposed in craters. However, Phoebe's most water-rich components are far redder than Nereid and have a less deep 4.5 \textmu m feature, suggesting that Nereid's spectrum is inconsistent with the capture and subsequent collisional processing of a water-rich KBO.}
	\label{fig:spectral_comparison} % give each figure a logical label name
\end{figure}

To examine the possibility that Nereid is a former member of the Kuiper belt that was subsequently captured into an irregular orbit around Neptune, we compare it to the JWST sample of KBO spectra. JWST has revealed three distinct spectral types in the Kuiper belt: water-rich, CO$_2$-rich, and methanol-rich, in order of hypothesized formation distance \cite{Pinilla-Alonso2025NatAs}. The methanol and CO$_2$-rich KBOs are substantially different from Nereid; these two spectral types have very red visible slopes, large aliphatic organic absorptions at 3.4~{\textmu}m, and CO ice. Nereid is better matched by water-rich KBOs (Fig.~\ref{fig:spectral_comparison}A), which are defined by a deep 3.0~{\textmu}m absorption accompanied by water ice bands at 1.5, 2.0, and 4.5~{\textmu}m, while lacking the numerous other volatiles seen on their CO$_2$ and methanol-rich counterparts. However, Nereid's H$_2$O ice spectral features are deeper than those on any observed water-rich KBO, while Nereid's 1.0 to 2.5~{\textmu}m slope is bluer than that of any KBO in the JWST sample. 

Nereid sees a higher impactor flux than KBOs \cite{Nesvorny2003AJ, Bottke2010AJ}, thus a better comparison to Nereid might be Phoebe, Saturn's heavily cratered irregular satellite thought to be a captured water-rich KBO \cite{Porco2005Sci, Fraser2018AJ, Belyakov2025PSJ}. Comparing Nereid with Phoebe's JWST global spectrum (Fig.~\ref{fig:spectral_comparison}B), Phoebe has a redder NIR slope and lower overall water ice abundance. Water-rich pixels in cratered regions of Phoebe from \textit{Cassini} VIMS fly-by data have a 2.0~{\textmu}m absorption band depth exceeding Nereid's (Fig.~\ref{fig:spectral_comparison}B), but have a lower albedo \cite{Porco2005Sci, Fraser2018AJ}, redder 1.0 to 2.5~{\textmu}m slope, bluer 2.2 to 3.5~{\textmu}m slope, and a shallower 4.5~{\textmu}m water ice absorption band than Nereid (Fig.~\ref{fig:spectral_comparison}B). Therefore, even when a water-rich KBO surface is processed to reveal deep water-ice absorptions, its spectrum does not become more Nereid-like, as demonstrated by the cratered regions on Phoebe being spectrally distinct from Nereid. Therefore, Nereid does not appear to be captured from the most common spectral type of dynamically excited KBOs.

Nereid's visible albedo is 0.24, higher than the 0.05-0.10 albedos of water-rich KBOs \cite{Pinilla-Alonso2025NatAs}, but similar to the least icy Saturnian or Uranian icy satellites, such as Hyperion or Umbriel \cite{Buratti1990Icar, Karkoschka2001Icar}. The closest match to Nereid's visible color and albedo is Hyperion \cite{Kiss2016MNRAS}, yet Hyperion has a spectrum even icier than the water-rich parts of Phoebe \cite{Clark2012Icar, BradDalton2012Icar, Brown2025PSJ}, with a more negative slope in the continuum across the 3.0~{\textmu}m water ice band. Water ice has a negative sloping continuum in the infrared \cite{Mastrapa2009ApJ}, therefore the darkening agent on Nereid's surface must be red in the 2.0 to 5.0~{\textmu}m region to set Nereid's flat continuum between 2.2 and 3.5~{\textmu}m. Organics and tholin-like materials are notably red in the visible and blue in the infrared unlike Nereid, while amorphous carbon is too dark with a flat 1--5~{\textmu}m slope \cite{Clark2012Icar}. Nano-phase hematite and iron have the Nereid-like blue visible and red infrared slope, however these silicate-derived materials usually display 1.0~{\textmu}m absorptions that are clearly seen on Phoebe or Hyperion \cite{Clark2012Icar} but are absent on Nereid. 

Another relevant parameter to our comparisons is Nereid's size. In the sample of KBOs observed with JWST, there are a dozen similarly-sized objects, including 1996 TL$_{66}$ and 2001 FP$_{185}$, both of which are water-rich KBOs that appear distinct from Nereid  \cite{Pinilla-Alonso2025NatAs}. Meanwhile, Haumea and its eponymous collisional family \cite{Brown2005ApJL} show much higher albedos than Nereid, and do not match its spectrum (Fig.~\ref{fig:spectral_comparison}B). While we cannot, of course, rule out the possibility of some Nereid-like object lurking unobserved within the Kuiper belt, the sample of observed KBOs cover objects from all major dynamical classes as well as an albedo, size, and color range that includes Nereid. For example, in the hot classical Kuiper belt, there are under a hundred objects in the absolute magnitude of Nereid (H$\sim3.5-5.5$, Nereid is 4.4) \cite{Petit2023ApJL}, of which JWST has sampled 10. 

Nereid's unique spectrum among outer solar system bodies is not consistent with a scenario where Nereid is captured during the early solar system's dynamical instability \cite{Nesvorny2007AJ}. The combination of a bright, water icy surface, with a unique dark material renders Nereid unlike KBOs or other irregular satellites like Phoebe, suggesting that Nereid did not form in the outer planetesimal disk alongside KBOs.

\subsubsection*{Nereid's dynamical origin}
As Nereid does not appear to be compositionally related to KBOs, two possibilities emerge for its origin. The first is capture from dynamically cold planetesimals local to Neptune, most of which were scattered or incorporated into the planet, and are therefore not well-represented in the present-day Kuiper belt \cite{Jewitt2007ARA&A}. This hypothesis presents multiple issues. Mechanisms of early, pre-instability capture are not efficient at the ice giants. Gas drag and capture by mass accretion of the host planet require large masses and extensive gas envelopes for which Neptune is too small \cite{Nesvorny2007AJ}. Binary capture provides an alternative mechanism \cite{Agnor2006Natur, Jewitt2007ARA&A}, however, objects captured too early have a high risk of disruption as large planetesimals and planetary bodies cross the giant planet region prior to and during the Solar System's early dynamical instability \cite{Nesvorny2007AJ}. Conversely, a post-instability capture is too late for the existence of a sizable reservoir of local planetesimals, and binary capture will then occur from more distant sources. 

Nereid's dynamical peculiarities raise the question of whether it originated as a regular satellite, perturbed by Triton to its current orbit \cite{Goldreich1989Sci}. We explore whether Nereid, starting as an initially regular satellite formed around Neptune, could have acquired its present orbit following the key event in the Neptunian system's history: Triton's capture and orbital evolution. Triton likely originated as a Pluto-like body with a large satellite, evinced by the similar compositions of Pluto and Triton \cite{McKinnon1984Natur}. During a close encounter with Neptune, the binary pair of Triton and its satellite was dissociated, thus capturing Triton on a highly-eccentric, retrograde orbit \cite{Agnor2006Natur}. From this initial state, Triton's orbit circularized to evolve onto its current tidally-locked orbit, maintaining its retrograde inclination. The paucity of information on the initial satellite system and the exact conditions of Triton's capture limits us to demonstrating the plausibility of Triton perturbing Nereid to an irregular satellite-like orbit, rather than determining a pathway and set of conditions that give an exact orbital match to present-day Nereid. 

Models of Triton's capture show that the dissociation of Triton from its binary initially produce a highly eccentric orbit, since the change in energy from a heliocentric to planetocentric orbit tends to leave Triton on a tenuous orbit around Neptune \cite{Agnor2006Natur, Nogueira2011Icar}. For Triton to remain bound to Neptune, it must rapidly lose orbital energy and circularize. While tides were originally suggested as the mechanism to bring Triton to its current orbit \cite{Nogueira2011Icar}, recent work has demonstrated that tidal evolution may be too slow to preserve Triton as it interacts with a disk of pre-existing satellites. Instead, circularization via dissipative collisions can proceed over an order of magnitude more rapidly than high-eccentricity tidal migration \cite{Cuk2005ApJL, Rufu2017AJ}. 

\begin{figure} % Do NOT use \begin{figure*}
	\centering
	\includegraphics[width=\textwidth]{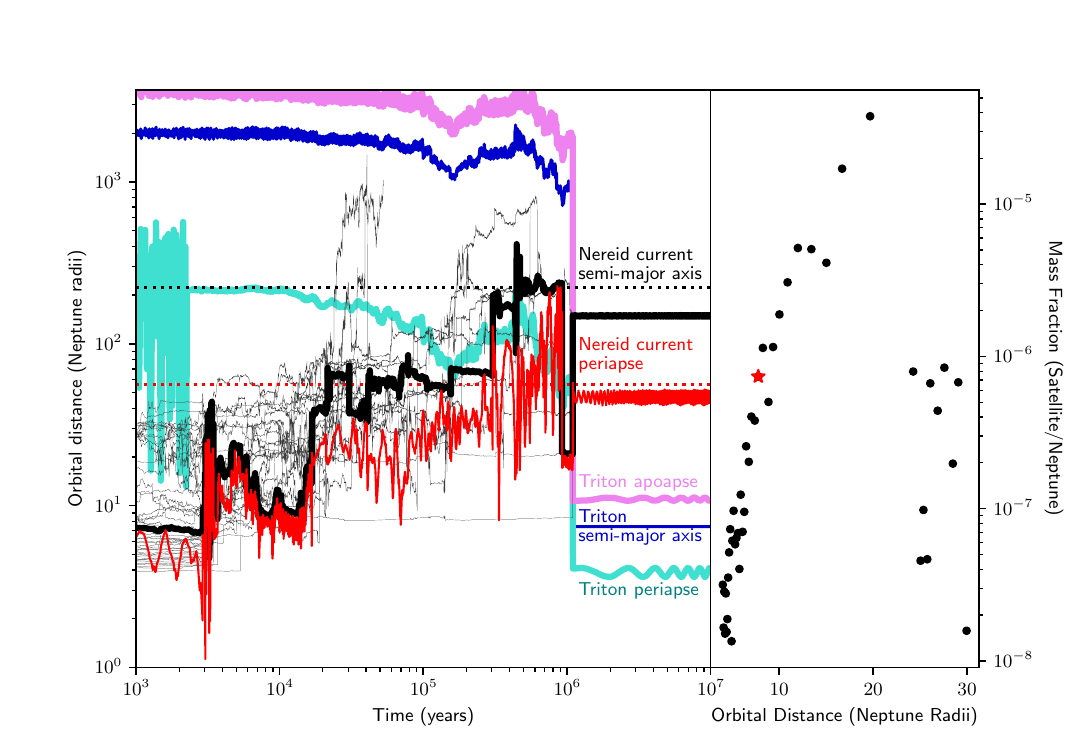} % for an image file named example_figure.*
	% Pick an appropriate width - in print, figures are usually one or two columns wide, which can
	% be approximated by 0.3\textwidth or 0.6\textwidth respectively. Use appropriate label sizes.

	% Captions go below figures
	\caption{\textbf{Evolution of Triton into a disk of satellites.} Left: Triton's semi-major axis, periapse, and apoapse are in blue, cyan, and violet respectively, while the satellite that lands on a Nereid-like orbit is shown with semi-major and periapse in black and red, with the semi-major axes of all other satellites shown as faint black lines. The present-day semi-major axis and periapse of Nereid are shown as black and red dotted lines, respectively. Triton, during its inwards migration, perturbs one of the initially regular satellites, and kicks it out to a Nereid-like orbit. The final inclination of the moon is 33 degrees with respect to Neptune's equator, similar to Nereid's inclination as computed relative to the local Laplace plane. Right: The initial configuration for the proto-Neptunian satellites is generated to match results for satellites generated by a vapor disk at Uranus \cite{Ida2020NatAs}. The red star marks the starting location and mass of the satellite that is perturbed to Nereid's orbit.}
	\label{fig:disk_triton_simulation} % give each figure a logical label name
\end{figure}

We simulate collisional circularization in order to determine the feasibility of transferring one of the pre-existing satellites to an irregular orbit as Triton disrupts the original Neptunian system. Our simulations include the Sun, Neptune, an eccentric and inclined Triton, and a satellite system with a Uranus-like mass at an evolved, but not final stage of satellite accretion. Evolving this system for a million years, we demonstrate that Triton is able to perturb initially regular satellites to an orbit within 10\% of Nereid's present semi-major axis and periapse, while clearing out most of the system and circularizing in the process (Fig. \ref{fig:disk_triton_simulation}). In the simulation shown, the final Nereid-like satellite has a $\sim$$40$ degree mean inclination relative to Neptune's equatorial plane and a mean eccentricity of 0.705. Nereid has a 28.4 degree inclination relative to Neptune's equator and an eccentricity of 0.75, thus our simulation can roughly reproduce Nereid's present-day orbit. 

Characterizing our entire suite of simulations, we see three kinds of outcomes. The most frequent result is the ejection or destruction of Triton, representing 60\% of runs, consistent with past work \cite{Rufu2017AJ}. Among simulations which do not leave Triton intact, nearly all produce one or more Nereid-like bodies. Considering the abundance of Triton-mass objects in our early Solar System \cite{Morbidelli2009Icar}, a plausible history for the Neptunian system and Nereid is that an early encounter with a Triton-sized body destroyed many of the original satellites while leaving Nereid on an irregular orbit. Present-day Triton is then a product of a second capture event, circularizing rapidly enough to preserve Nereid even if most of the system is gone. In fact, such a scenario is favorable for Triton's capture, as previous work indicates that a system mass that is 20\% smaller than that of the Uranian moons is more likely to capture Triton than a Uranian-like system \cite{Rufu2017AJ}. The second outcome of our simulations, at 20\% of runs, is one where Triton removes all the satellites except a set of inner moons, leaving the system much like the present-day one, but without Nereid. Finally, we have 20\% of strictly successful simulations, in which Triton circularizes and leaves one or more satellites on irregular orbits. We define any object with semi-major axis greater than R$_N>50$, and either an inclination greater than 10 degrees or eccentricity more than 0.2 as irregular satellite like. Were a satellite with such characteristics observed, it would evoke the same questions raised by Nereid. The probability of Triton both surviving an encounter with a Uranus-like system and kicking a regular satellite to a dynamically stable irregular orbit is thus approximately 20\%. We show the range of outcomes described above in the supplementary material \cite{supplmat}.

\subsection*{Discussion}
Nearly 70 years after Nereid's discovery by Kuiper and 40 years after the brief Voyager 2 fly-by, we have demonstrated that Nereid's spectroscopic characteristics do not appear to be consistent with irregular satellite capture \cite{Nesvorny2007AJ}, as its near-infrared spectrum does not resemble that of any KBO or irregular satellite observed with JWST to date. Our proposed regular satellite genesis story for the moon leaves Nereid as the singular intact original satellite of Neptune --- Neptune's innermost moons, such as Proteus, are re-accreted pieces of satellites destroyed by Triton's capture \cite{Banfield1992Icar, Rufu2017AJ}. Further details in Nereid's spectrum not revealed by our study, such as the deuterium-to-hydrogen ratio, carbon isotope ratios, or precise CO$_2$ band center could unlock answers to key questions regarding the evolution of Neptune and the characteristics of its original satellite system. Future spacecraft exploration of the Neptunian system should search for signs of an early geologic history on Nereid consistent with formation as a regular satellite, as tidal evolution of Triton over billions of years has erased its collisional past \cite{mousis_triton}. Additionally, finding evidence for the collisional capture of Triton through observations of inner moons such as Proteus and Larissa would help confirm our dynamical story for the Neptunian system. 

The destruction of the initial Neptunian satellite system precludes us from definitively tracing Nereid's history from grains to accretion to transfer to an irregular orbit. Reliable inferences of the probability of the dynamical events described occurring would require knowledge of the mass of Nereid, the exact configuration of satellites in Neptune's primordial system, and a constraint on the timing of Triton's capture. Our outlined scenario for Nereid favors a mass for the original Neptunian satellite system that was either smaller than that of the Uranian system, or that the mass became smaller through interactions with large interlopers prior to Triton's capture. Such a system would both facilitate Triton's capture and have more mid-sized satellites than the present Uranian system, increasing the number of Nereid-like bodies that could be perturbed. Additionally, the collisional model for Triton's capture prefers a capture for Triton shortly after the formation of the Neptunian satellites, before they had time to evolve into a version of the present Uranian system. Perturbing a regular satellite to a Nereid-like orbit or a more distant one depends on the circularization timescale, where a longer circularization for Triton leaves more time for satellites to diffuse outwards, causing them to be kicked to yet more distant orbits. The possible parameter space for the initial configuration of the Neptunian system and Triton's capture is effectively unconstrained and is far larger than can be explored through simulations. As such, our simulations are merely illustrative of the process wherein a smaller body undergoing scattering is stabilized by interactions with a massive external perturber, demonstrating that a transfer of Nereid from a circular to eccentric and inclined orbit in the context of Triton's circularization is plausible. We note that the process we have outlined, though on much grander scales, has recently been shown to deliver planets to distant, yet bound orbits during dynamical instabilities in the early Solar System \cite{Izidoro2025NatAs}. 

We note that there exists another mechanism for exciting regular satellites to Nereid-like orbits that has been proposed in the literature on giant planet migration, supporting the dynamical scenario we have outlined. The early Solar System contained hundreds of Triton-sized bodies which were ejected by scattering off of the giant planets; this process is partially responsible for the migration of the giant planets \cite{Fernandez1984Icar, Tsiganis2005Natur}. Simulations of close encounters between Neptune and $>$$1000$km bodies demonstrate that regular satellites past 12 Neptune radii are readily excited to Nereid-like orbits \cite{Beauge2002Icar}. More severe planet-planet close encounters can produce similar results, although planet-planet scattering events require fine-tuned initial conditions to produce Nereid without removing the entire proto-Neptunian system \cite{Li2020AJ}. 

Finally, we discuss the effectiveness of dynamical models for capture of irregular satellites from the Kuiper belt, as applied to Nereid. Irregular satellite capture that proceeds from planet-planet interactions perturbing KBOs onto planetocentric orbits appears to be rather inefficient at delivering Nereid from the Kuiper belt to its present orbit \cite{Nesvorny2007AJ, Nesvorny2014ApJ}. The original number of Nereid-sized objects in the proto-Kuiper belt is at most of order $10^5$ bodies \cite{Morbidelli2009Icar, Schlichting2011ApJ, Kenyon2012AJ}. Capture efficiency for irregular satellites at Neptune is around $10^{-8}$ \cite{Nesvorny2014ApJ}. Thus, the probability of capturing a single Nereid-like body from the proto-Kuiper belt is close to 0.1\%, far lower than in our regular satellite origin story. An alternative capture regime is binary capture, as discussed for Triton. Capturing Nereid as the primary member of a binary is rather unlikely, as shown in three-body capture simulations \cite{Philpott2010Icar,Gaspar2013MNRAS}. If Nereid were instead the smaller member of a binary, capture is possible only for wide binary separations, at least as wide as Pluto/Hydra \cite{Agnor2006Natur, Kobayashi2012ApJ}. While we cannot rule out that Nereid was captured from a binary KBO, the numerous Pluto-mass objects of the early Solar System would have had their moons dispersed into the Kuiper belt, and would appear in the present-day record of 400km-sized bodies \cite{Nesvorny2019Icar}. A regular satellite origin for Nereid as proposed in our work and hinted at in previous studies \cite{Beauge2002Icar, Li2020AJ} appears more likely than a rare capture of an object with no known spectral analogues in the present-day Kuiper belt.

\subsection*{Materials and Methods}
\subsubsection*{JWST data reduction}
Nereid was observed with JWST NIRSpec on November 23, 2024 between 03:28:53 UT and 05:24:45 UT as part of the Cycle 3 GO Program \#4645. The observations were obtained using the Integral field unit (IFU) in PRISM mode with a nominal R$\sim$30-300 resolving power over a 0.6-5.3~{\textmu}m wavelength range. We used a four point dither configuration with a 160.5 second exposure time per dither. All other JWST spectra presented in this study were obtained from the Mikulski Archive for Space Telescope (MAST), and can be found at \url{https://doi.org/10.17909/2bj1-h545}. The spectrum of Nereid and all other objects shown were reduced using the PSF template fitting method that has been applied to numerous studies with JWST \cite{Brown2023PSJ, Wong2024PSJ}. The reduction code, \texttt{jwstspec}, can be found online on Zenodo \cite{jwstspec}.

\subsubsection*{Dynamical simulations}
All of our simulations use the~\textsc{REBOUND} N-body integrator \cite{Rein2012A&A}, and the code used to initialize all of our simulation runs is available for download at CaltechData (\url{https://doi.org/10.22002/8t37m-qfr21}) \cite{caltechdata}. Our simulations include the Sun, Neptune, Triton, and a disk of massive satellites. We use the adaptive timestep \textsc{IAS15} integrator to ensure high precision during close encounters. We initialize Triton at a periapse of 14 or 21 Neptune radii, and vary semi-major axis distance from 1300 to 2300 R$_N$. Triton starts with its present-day inclination and randomized longitudes. We model the satellite system of Neptune to mimic the distribution of satellite masses and semi-major axes produced by the vapor disk model for Uranian satellite formation \cite{Ida2020NatAs}. We generate moons between 3 and 20 Neptune radii following the isolation mass for oligarchic growth \cite{Ida2020NatAs}, which is given by:
\begin{equation}
    m_{\text{isolation}} = 0.74 \times 10^{-4} \left(\frac{r}{20r_{\text{Nep}}}\right)^{21/4}M_{\text{Nep}} 
\end{equation}
This distribution of satelitesimals would be replicated by other models of satellite formation, such as those from tidal disruption of comets \cite{Charnoz2010Natur, Hyodo2017Icar}. We set the total mass of the disk to $10^{-4}\times M_{\text{Nep}}$, generating a system similar in mass to the Uranian system. We also run simulations with a smaller system mass, at $2 \cdot 10^{-5}\times M_{\text{Nep}}$. We cap the largest generated satellites at $1\times10^{-5}$ $M_\text{Neptune}$, and multiply their masses by a random factor between 0.25 and 4. Past the cut-off of 20 $r_{\text{Nep}}$, we use a log-normal distribution centered at $10^{-7}$ $M_\text{Nep}$ to generate satellites out to 30 $r_{\text{Nep}}$ or until we reach our mass cap. Our resulting model satellite disk represents an intermediate stage of oligarchic growth of satellites at Neptune. A key point in our simulations is that we handle collisions by merging particles while conserving mass, momentum, and energy, and track these collisions in order to precisely ascertain which original moons survive the encounter.

% Referring to supplementary material:
% Whenever more details are given in the Materials and Methods section, cite an entry in
% the reference list that directs readers there, like this~\cite{methods}.
% To refer to material in the Supplementary Text section, just write (Supplementary Text).
% See guidance below for the difference between those two types of supplementary material.
% Supplementary figures and tables are referred to in lowercase
% e.g.~figure~\ref{fig:sup_example} or table~\ref{tab:sup_example}.
% Material in separate files needs to be hand coded e.g.~data~S1, movie~S2.

\newpage

%%%%%%%%%%%%%%%% MAIN TEXT FIGURES %%%%%%%%%%%%%%%

%%%%%%%%%%%%%%%% MAIN TEXT TABLES %%%%%%%%%%%%%%%

% \begin{table} % Do NOT use \begin{table*}
% 	\centering
% 	% Captions go above tables
% 	\caption{\textbf{All captions must start with a short bold sentence, acting as a title.}
% 		Then explain what is being listed in the table, the meaning of each column etc.
% 		Captions are placed above tables.}
% 	\label{tab:example} % give each table a logical label name
	
% 	\begin{tabular}{lccc} % four columns, alignment for each
% 		\\
% 		\hline
% 		Sample & $A$ & $B$ & $C$\\
% 		 & (unit) & (unit) & (unit)\\
% 		\hline
% 		First & 1 & 2 & 3\\
% 		Second & 4 & 6 & 8\\
% 		Third & 5 & 7 & 9\\
% 		\hline
% 	\end{tabular}
% \end{table}

%%%%%%%%%%%%%%%% REFERENCES %%%%%%%%%%%%%%%

\clearpage % Clear all remaining figures and tables then start a new page

% The list of references goes after the main text and before the acknowledgements
% When preparing an initial submission, we recommend you use BibTeX, like this:
%
\bibliography{nereid} % for a file named science_template.bib
\bibliographystyle{sciencemag}

% After the paper has completed peer review and been revised ready for acceptance,
% you should comment out the lines above and copy-paste the contents of your .bbl
% file here instead. This will help ensure that our conversion software works correctly.
% Remember to re-run BibTeX first - check the timestamp!
%
% Example of the first three entries copy-pasted from science_template.bbl:
%
%\begin{thebibliography}{1}
%
%\bibitem{example}
%A.~N. {Author}, An example reference. \emph{Journal of Improbable Research}
%  \textbf{1}, 67 (2020).
%
%\bibitem{example2}
%F.~M. {Surname}, S.~{Author}, A second example. \emph{Interesting Research
%  Letters} \textbf{32}, 897 (2019).
%
%\bibitem{example_preprint}
%P.~{One}, P.~{Two}, P.~{Three}, {An unpublished preprint}. \emph{preprint}
%  (2021), arXiv:2101.12345.
%
%\end{thebibliography}

%%%%%%%%%%%%%%%% ACKNOWLEDGEMENTS %%%%%%%%%%%%%%%

\section*{Acknowledgments}
This study was based on observations made with the NASA/ESA/CSA James Webb Space Telescope (JWST). The data were obtained from the Mikulski Archive for Space Telescopes (MAST) at the Space Telescope Science Institute, which is operated by the Association of Universities for Research in Astronomy under NASA contract NAS 5-03127 for JWST. The authors thank Wes Fraser for his contribution of Cassini VIMS data on Phoebe, Marina Brozovic for providing accurate orbital elements of Nereid, Tiger Lu for assistance with the dynamical simulations, Samantha Trumbo for valuable discussions, and Zachariah Milby for plot styling.
\paragraph*{Funding:}
The observations analyzed here are associated with JWST program \#4645. Support for program \#4645 was provided by NASA through a grant from the Space Telescope Science Institute, which is operated by the Association of Universities for Research in Astronomy, Inc., under NASA contract NAS 5-03127.
\paragraph*{Author contributions:}
M.B. performed the analysis, led the interpretation, and wrote the manuscript. M.R.D. contributed to the interpretation of the spectrum and aided in manuscript writing. I.W. reduced the JWST Nereid data. K.B. contributed to the interpretation of the dynamics. M.E.B. reduced the JWST Kuiper belt and Phoebe data, contributed to the interpretation, and aided in manuscript writing.
\paragraph*{Competing interests:}
The authors declare that they have no competing interests.
\paragraph*{Data and materials availability:}
All data and code needed to evaluate and reproduce the results in the paper are present in the paper and/or the Supplementary Materials. The JWST data are available at MAST: (\url{https://mast.stsci.edu/portal/Mashup/Clients/Mast/Portal.html}) under program ID 4645 for Nereid, program ID 3716 for Phoebe and Hyperion, and program IDs 1191, 1272, and 2418 for the Kuiper belt object spectra, with the specific observations used stored at: \url{https://doi.org/10.17909/2bj1-h545}. We have also used data from proposal ID 1128 for calibration with a solar-like spectrum, however use of any available solar analogue spectrum is sufficient. The VIMS spectra of Hyperion and Phoebe were obtained from past work by digitization and author request (see refs \cite{Clark2012Icar,Fraser2018AJ}). The code to plot the spectra and the orbital simulations are both available at CaltechData (\url{https://data.caltech.edu/records/8t37m-qfr21}) \cite{caltechdata}. This study did not generate new materials.

% Specify where the data, software, physical samples, simulation outputs or other materials
% underlying the paper are archived.
% They must be publicly accessible when the paper is published (without embargo) and enable
% readers to reproduce all the results in the paper.
% Contact the editor if you’re unsure what needs to be shared.
% Our preference is for digital material to be deposited in a suitable non-profit online data or
% software repository that issues the material with a DOI.
% Alternatively, an institutional repository, subject-based archive, commercial repository etc.
% is acceptable, as are (short) supplementary tables or a machine-readable supplementary data file.
% ‘Available on request’ or personal web pages are not allowed.

% Cite the relevant DOI \cite{dataset}, URL \cite{example_url} or reference \cite{example2}
% in this statement.
% These \textit{do not} count towards the reference limit if they are only cited in the acknowledgements.
% Be specific and state a unique identifier -- such as an accession number, software version number
% or observation ID -- so readers can easily retrieve the exact material used.

%%%%%%%%%%%%%%%% SUPPLEMENT LIST %%%%%%%%%%%%%%%

% List the contents of your Supplementary Materials, including the numbers of any
% supplementary figures, tables, external data files etc. and any references that are
% cited only in the supplement. In this example, refs. 7-8 are cited only in the supplement.
% Fill out your numbers accordingly and delete any lines that aren't applicable.
\subsection*{Supplementary materials}
Supplementary Text\\
Figures S1-S3\\

%%%%%%%%%%%%%%%% END OF MAIN TEXT %%%%%%%%%%%%%%%

\newpage

%%%%%%%%%%%%%%%% START OF SUPPLEMENT %%%%%%%%%%%%%%%

% Figures, tables, equations and pages in the supplement are numbered S1, S2 etc.
\renewcommand{\thefigure}{S\arabic{figure}}
\renewcommand{\thetable}{S\arabic{table}}
\renewcommand{\theequation}{S\arabic{equation}}
\renewcommand{\thepage}{S\arabic{page}}
\setcounter{figure}{0}
\setcounter{table}{0}
\setcounter{equation}{0}
\setcounter{page}{1} % not 0 as \newpage already started a supplementary page
% References continue the numbering from the main text.

%%%%%%%%%%%%%%%% SUPPLEMENT TITLE PAGE %%%%%%%%%%%%%%%

\begin{center}
\section*{Supplementary Materials for\\ \scititle}

% Author list for the supplement
% Indicate the corresponding authors, but do NOT include institutions here
% It would be nice if the template auto-generated this, but doing so is complicated...
Matthew~Belyakov$^{\ast}$,
M.Ryleigh~Davis,
Ian~Wong,
Konstantin~Batygin,\\
Michael~E.~Brown\\
\small$^\ast$Corresponding author. Email: mattbel@caltech.edu\\
\end{center}

% Fill out the numbers for each type of supplementary material,
% and delete any lines that aren't applicable.
% These are just example numbers that don't match the rest of this template.
\subsubsection*{This PDF file includes:}
Supplementary Text\\
Fig. S1 and S2\\

\newpage

\subsection*{Other Simulation Outcomes}
\begin{figure}
    \centering
    \includegraphics[width=1\linewidth]{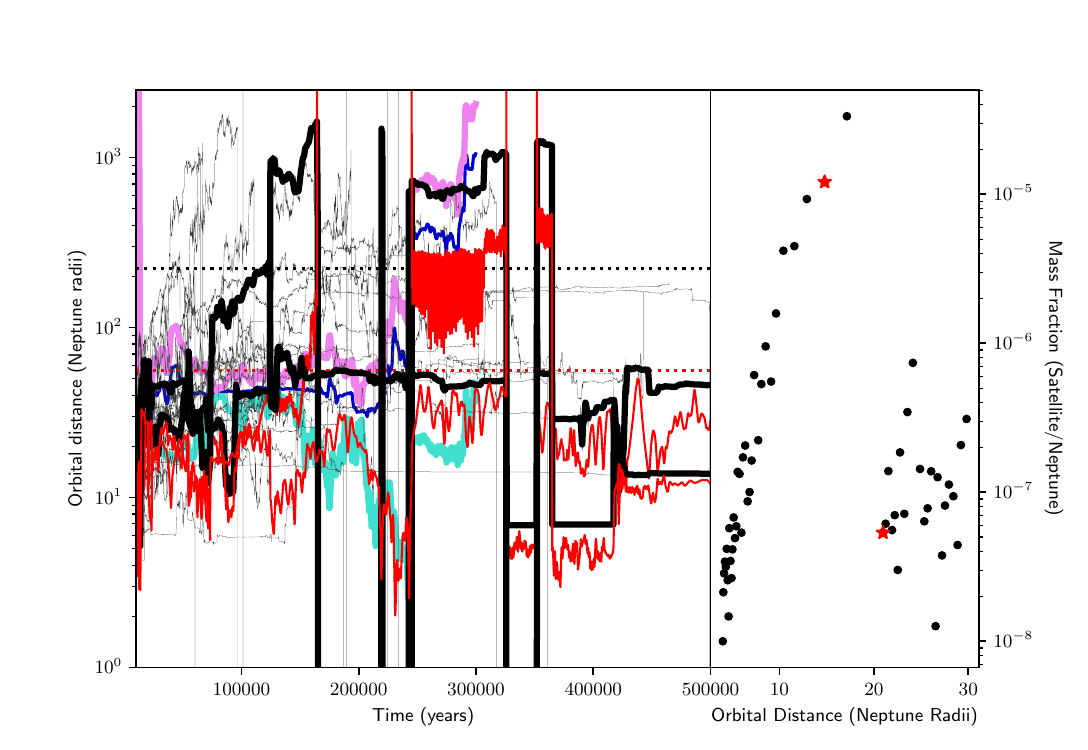}
    \caption{An example of a simulation where Triton is ejected from the system, leaving behind one satellite which lands on a low-eccentricity orbit, while one has significant inclination and eccentricity.}
    \label{fig:no_triton}
\end{figure}

\begin{figure}[h]
    \centering
    \includegraphics[width=1\linewidth]{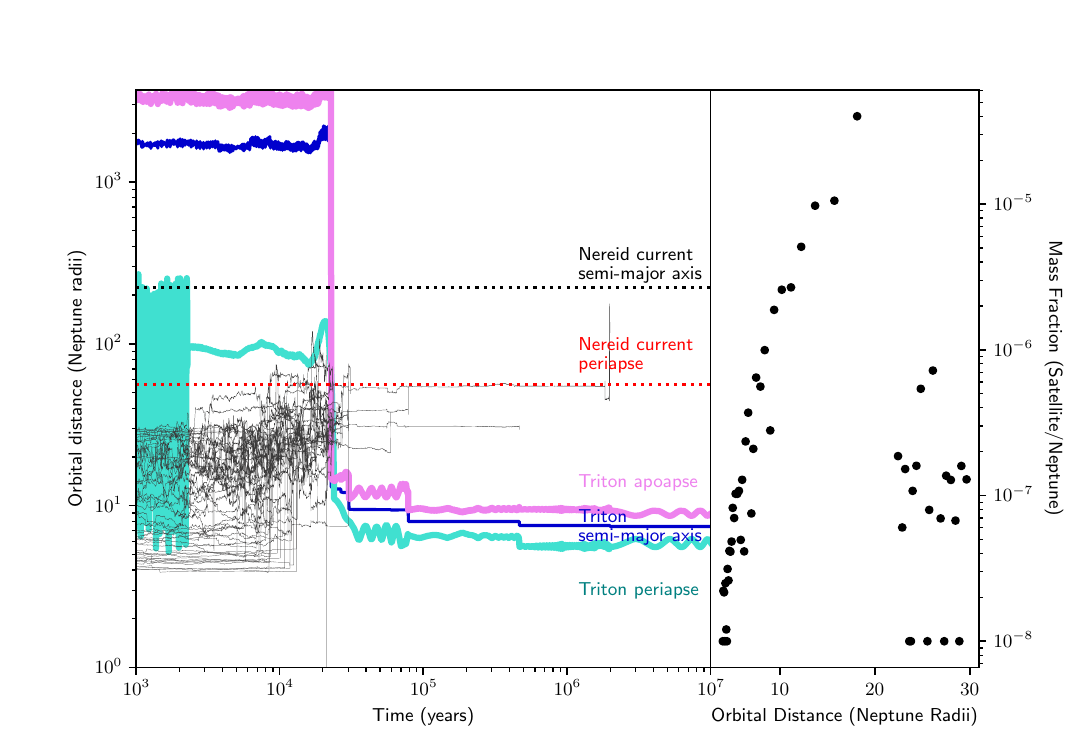}
    \caption{An example of an unsuccessful simulation run, where Triton circularizes and all of the satellites get ejected, or collide with each other, Triton, or Neptune, with the latter being the most common outcome.}
    \label{fig:disk_fail}
\end{figure}

\begin{figure}
    \centering
    \includegraphics[width=1\linewidth]{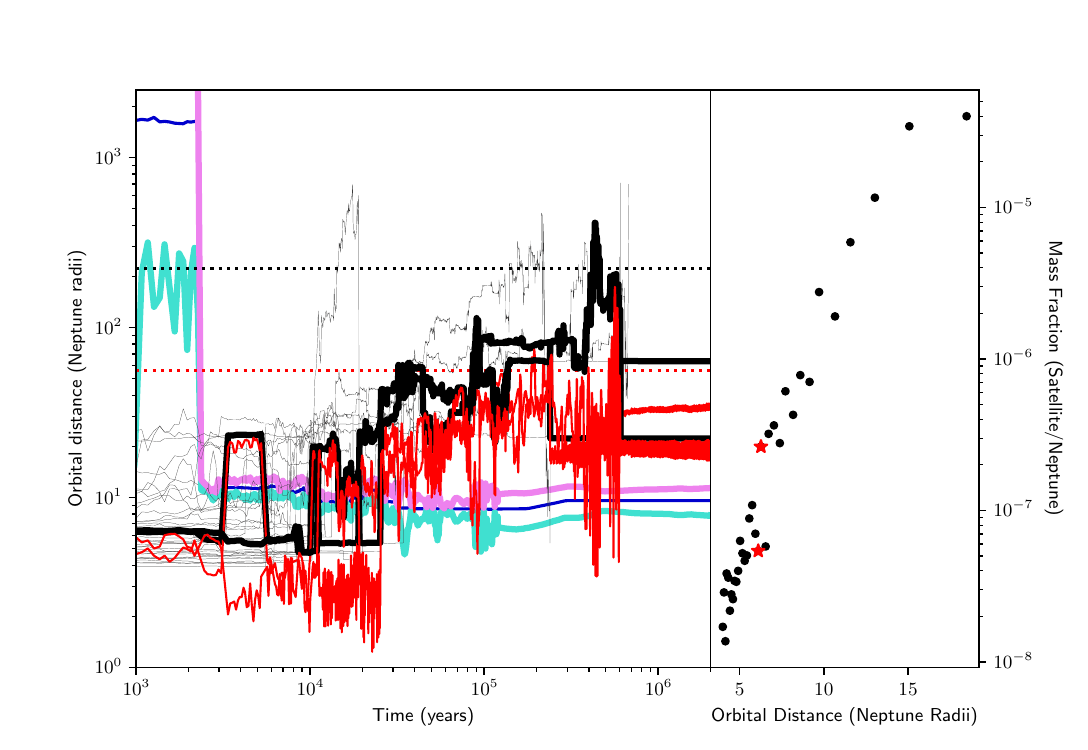}
    \caption{An example of a simulation where three objects successfully cross Triton's apoapse, one of which lands on a relatively low-eccentricity orbits, while one is closer to Nereid's orbit.}
    \label{fig:disk_weird}
\end{figure}

In the supplementary material, we show other simulation outcomes besides the one shown in Figure 3 of the main text, using the same plotting scheme. The most common outcome of our simulations as shown in Fig. \ref{fig:no_triton} is Triton being ejected (or colliding with Neptune) similar to other works that have simulated the same process \cite{Rufu2017AJ}. Typically, this process leaves several irregular satellites in orbit. In Fig. \ref{fig:disk_fail} we show an example of a simulation run where Triton circularizes and destroys the satellite system without having perturbed a moon to a Nereid-like orbit. We also observe simulations in which Triton perturbs several satellites to irregular orbits (Fig. \ref{fig:disk_weird}). Occasionally, these moons also happen to acquire retrograde or relatively circular and distant orbits -- irregular, though not necessarily Nereid-like.

\end{document}